\documentclass[sigchi, authorversion]{acmart}
\usepackage{svg}
\usepackage{algorithm2e}
\usepackage{graphics}
\usepackage{float}
\usepackage{wrapfig}
\usepackage{array}
\usepackage{makecell}
\usepackage[inline]{enumitem}
\usepackage{listings}
\usepackage{xcolor}
\usepackage{url}

\AtBeginDocument{%
  \providecommand\BibTeX{{%
    \normalfont B\kern-0.5em{\scshape i\kern-0.25em b}\kern-0.8em\TeX}}}

\acmYear{2024}
\setcopyright{cc}
\acmConference[CHI '24]{Proceedings of the CHI Conference on Human Factors
in Computing Systems}{May 11--16, 2024}{Honolulu, HI, USA}
\acmBooktitle{Proceedings of the CHI Conference on Human Factors in
Computing Systems (CHI '24), May 11--16, 2024, Honolulu, HI, USA}
\acmDOI{10.1145/3613904.3641915}
\acmISBN{979-8-4007-0330-0/24/05}

\makeatletter
\newcommand{\subsubsubsection}{\@startsection{paragraph}{4}{0ex}%
   {-3.25ex plus -1ex minus -0.2ex}%
   {1.5ex plus 0.2ex}%
   {\normalfont\normalsize\bfseries}}
\makeatother

\newcommand{\oldtexttt}\texttt

\definecolor{gptpurple}{RGB}{217,210,233}
\definecolor{gptgreen}{RGB}{217,234,211}
\definecolor{gptgray}{RGB}{230,230,230}
\definecolor{gptbeige}{RGB}{255,248,229}
\definecolor{gptred}{RGB}{244,204,204}

\definecolor{Fuchsia}{HTML}{8C368C}
\definecolor{ForestGreen}{HTML}{009B55}
\definecolor{Tan}{HTML}{DA9D76}
\definecolor{BrickRed}{HTML}{B6321C}

\def\gptbox #1 {\textbf{\textcolor{Fuchsia}{#1}} \hspace{0.3mm}}
\def\greengptbox #1 {\textbf{\textcolor{ForestGreen}{#1}} \hspace{0.3mm}}
\def\beigegptbox #1 {\textbf{\textcolor{Tan}{#1}} \hspace{0.3mm}}
\def\redgptbox #1 {\textbf{\textcolor{BrickRed}{#1}} \hspace{0.3mm}}
\def\greenbox #1 {\textbf{\textcolor{ForestGreen}{#1}} \hspace{0.3mm}}
\def\beigebox #1 {\textbf{\textcolor{Tan}{#1}} \hspace{0.3mm}}
\def\redbox #1 {\textbf{\textcolor{BrickRed}{#1}} \hspace{0.3mm}}
\newcommand{\GobbleListArg}[1]{, #1\CheckListArg}
\newcommand{\CheckListArg}{\csname @ifnextchar\endcsname\bgroup{\GobbleListArg}{\texttt{)}\egroup}}

\newcommand{\killpunct}[1]{}

\begin{document}

\title{VAL: Interactive Task Learning with GPT Dialog Parsing}

\author{Lane Lawley}
\affiliation{%
  \institution{Georgia Institute of Technology}
  \streetaddress{85 5th St. NW}
  \city{Atlanta}
  \state{Georgia}
  \country{USA}
  \postcode{30332}
}
\email{lanelawley@gmail.com}
\orcid{0000-0002-8961-6789}

\author{Christopher J. MacLellan}
\affiliation{%
  \institution{Georgia Institute of Technology}
  \streetaddress{85 5th St. NW}
  \city{Atlanta}
  \state{Georgia}
  \country{USA}
  \postcode{30332}
}
\email{cmaclell@gatech.edu}
\orcid{0000-0003-3084-5189}

\renewcommand{\shortauthors}{Lawley and MacLellan}

\begin{abstract}

Machine learning often requires millions of examples to produce static, black-box models. In contrast, interactive task learning (ITL) emphasizes incremental knowledge acquisition from limited instruction provided by humans in modalities such as natural language. However, ITL systems often suffer from brittle, error-prone language parsing, which limits their usability. Large language models (LLMs) are resistant to brittleness but are not interpretable and cannot learn incrementally. We present VAL, an ITL system with a new philosophy for LLM/symbolic integration. By using LLMs only for specific tasks—such as predicate and argument selection—within an algorithmic framework, VAL reaps the benefits of LLMs to support interactive learning of hierarchical task knowledge from natural language. Acquired knowledge is human interpretable and generalizes to support execution of novel tasks without additional training. We studied users' interactions with VAL in a video game setting, finding that most users could successfully teach VAL using language they felt was natural.

\end{abstract}



\begin{teaserfigure}
\centering
  \includegraphics[width=0.8\textwidth]{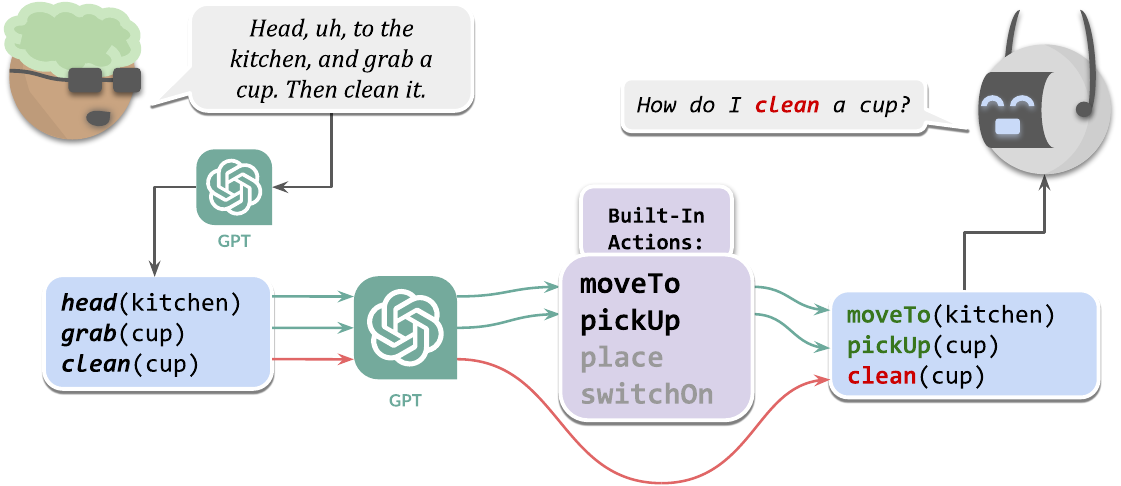}
  \caption{An illustration of one iteration of the basic VAL learning procedure. A more comprehensive process description is given by Algorithm~\ref{alg:valgorithm} and Section~\ref{sec:gpt_subroutine_arch}.}
  \Description{A cartoonish flowchart illustration of a human being instructing the VAL system, which asks a clarifying question after processing the input through several targeted GPT prompts.}
  \label{fig:teaser}
\end{teaserfigure}


\maketitle

\section{Introduction}
The Interactive Task Learning (ITL) approach introduced by \citet{itl} articulates a research vision dedicated to the goal of enabling machines to interactively learn general tasks by interacting with humans in natural ways. It posits that acquired task knowledge should be general enough to be applicable in novel situations, and that it should be interpretable and modifiable, so that human teachers---including non-technical end users---can actively shape the machine's understanding. 
ITL highlights many possible instructional modalities for the acquisition of task knowledge, including gestures, diagrams, and language---these modalities are also formally accounted for in the Natural Training Interaction (NTI) framework due to \citet{nti}.
Realizing the ITL vision requires interdisciplinary research that combines new machine learning techniques (e.g., to learn effectively from the limited instruction a single end user can provide) with user studies investigating what makes teaching interactions both natural and productive. 
In this work, we focus on the language modality, endeavoring toward acquisition of human-interpetable, modular, hierarchical task knowledge from natural dialogs between humans and machines.

\begin{figure}
    \begin{center}
    \includegraphics[width=\columnwidth]{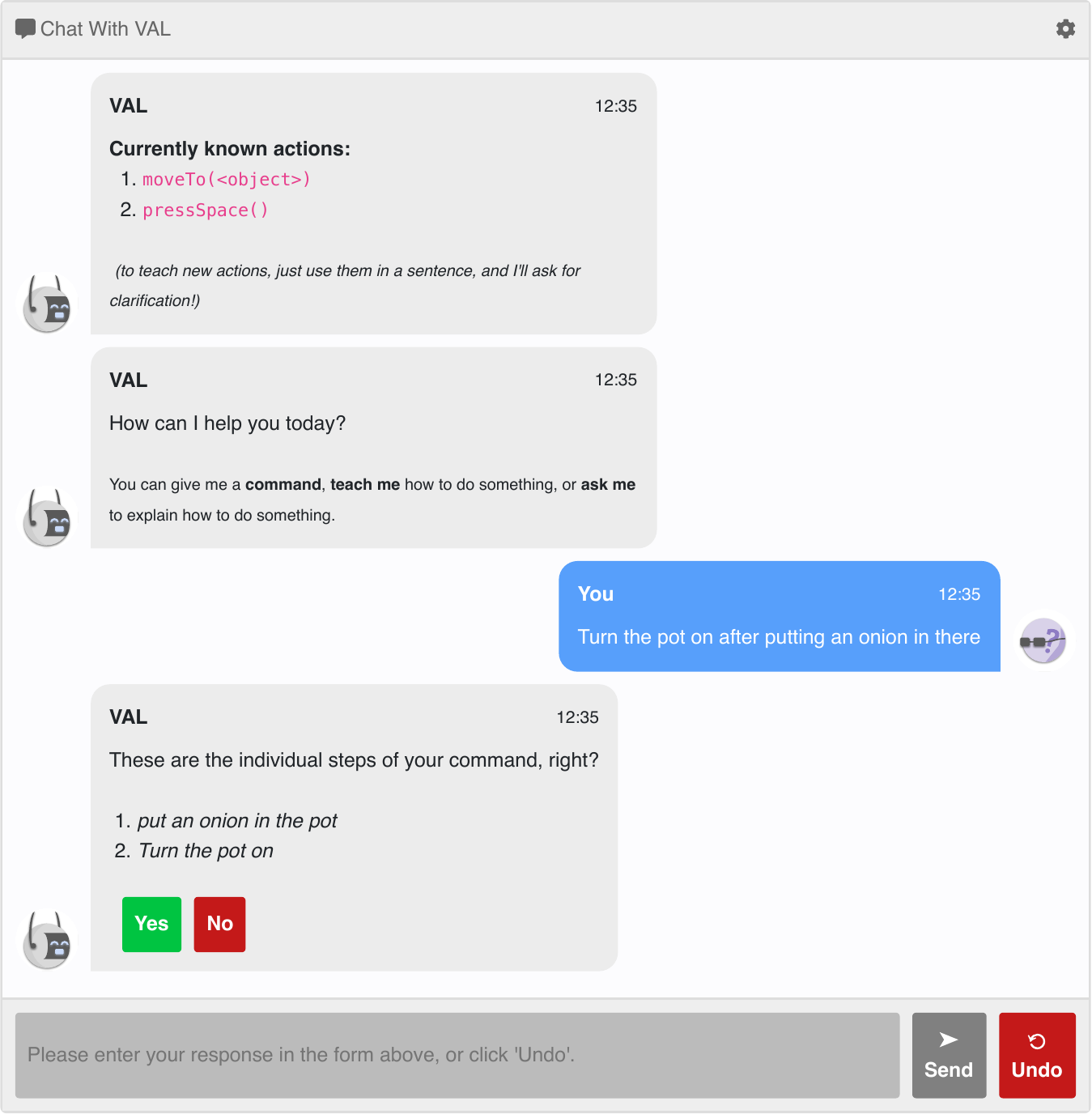}
    \end{center}
    \caption{An example dialog with VAL. The current dialog state is a confirmatory prompt for the text segmentation step performed by one of VAL's GPT subroutines. This step performs action discretization, anaphora resolution (of ``in there''), and temporal ordering.}
    \Description{An example dialog with VAL. The user asks VAL to ``turn the pot on after putting an onion in there'', and VAL responds with the broken down steps of the command.}
    \label{fig:val_dialog}
\end{figure}


There has been considerable research into language-driven task acquisition, which we detail further in Section~\ref{sec:background}. Each of these systems, however, suffers the characteristic brittleness of the classical syntactic and semantic parsers they invariably use to convert natural instructional dialogs into task knowledge. Syntactic forms vastly outnumber their semantic counterparts, and problems like anaphora resolution, paraphrasing, and grammatical and transcription errors frequently claim the ``natural'' in ``natural language'' as their first casualty. These systems not only fall subject to myriad parser errors on malformed input, but also lack the ability to perform adequate paraphrase resolution when parsed verb structures do not match known ones exactly, generally relying on manually constructed knowledge bases of synonyms to solve even a portion of this problem.
These issues make it unlikely that these systems will successfully achieve the ITL vision with real users---currently, they only work with users in the lab that know their idiosyncratic syntax.

Here, we present \textit{VAL}, the Verbal Apprentice Learner, an alternate strategy for mitigating the sheer variety of natural language. By exploiting the virtual mastery of the \textit{form} of natural English apparently acquired by large language models \citep{bert-rediscovers-nlp}, we demonstrate the acquisition of task knowledge in a far less restrictive, far more natural dialog setting than possible in prior work. We integrate a GPT-family language model \citep{gpt-3,rlhf,gpt-4} in a careful and principled way, using it to map natural dialog into a symbolic domain.
The GPT model is used for one of a limited number of sub-tasks: the semantic parsing of text into predicate-argument structures, the semantic unification of those structures to already-known actions with the same meaning, and other narrowly scoped procedures enumerated in Section~\ref{sec:gpt_subroutine_arch}.
The linguistic flexibility provided by GPT allows VAL to deal with error-laden and grammatically unrestricted input dialog, and to perform paraphrasal mappings that take into account meaning; together, these applications of GPT vastly widen the set of possible inputs to the task learning system with minimal upfront engineering cost.

We feel VAL is uniquely suitable for task knowledge acquisition: unlike LLMs alone, VAL can learn long-term, re-usable task structures with just a few examples; and unlike hierarchical task networks alone, VAL offers a rich natural language interface to guide task execution and learning. We provide evidence of VAL's usability collected from user studies, analyzing both subjective responses and objective signals.

\subsection{Key Contributions}
This paper makes the following key contributions:

\begin{enumerate}
    \item \textbf{A neuro-symbolic hybrid approach to interactive task learning} (Section~\ref{sec:design}) in which:
    \begin{enumerate}
        \item a classical learning algorithm is augmented with subroutines implemented as highly specific GPT prompts, and;
        \item unknown actions are recursively reduced to known ones via natural dialog.
    \end{enumerate}
    \item \textbf{The VAL system} (Section~\ref{sec:system}), an implementation of this approach with an emphasis on mitigating ``cascading errors'' through confirmations and undo operations.
    \item \textbf{A user study} (Section~\ref{sec:study}) of VAL's usability and efficacy in a video game task learning environment, evaluating both user-reported experience and objective measures of efficacy.
\end{enumerate}


\section{Background}
\label{sec:background}

\lstset{language=Lisp,
    basicstyle=\ttfamily\footnotesize,
    stringstyle=\ttfamily\color{black},
    commentstyle=\ttfamily\color{black},
    keywordstyle=\ttfamily\color{black},
    identifierstyle=\texttt,
    tabsize=2
    }
\begin{figure}
\begin{lstlisting}[escapechar=ß,breaklines=true]
M: The game is tic-tac-toe.
ß\textbf{R: I don't know that game, how many players are there?}ß
M: Two.
ß\textbf{R: Please start by teaching me the name of a legal action in the game.}ß
M: Place
ß\textbf{R: What are the verb and parameter arguments associated with this action?}ß
M: move 1 to 2
\end{lstlisting}

\caption{A conversation with the Rosie system \citep{rosie2} demonstrating the rigid nature of the interaction. \textbf{M} denotes a user message, and \textbf{R} denotes a message from Rosie.}
\Description{A conversation with Rosie, another task learning system. The user and Rosie go back and forth using less-than-natural language.}
\label{fig:rosie_chat}
\end{figure}

The ITL framework builds on and extends prior HCI research on programming by demonstration \citep{cypher1993watch,lieberman2001your,mcdaniel1999getting}, investigating a broader range of interactions and modalities (beyond demonstrational) that let users transfer knowledge into AI systems in a easy, scalable, and performant way.
As \citet{prog_users_too} note, even ``Programmers Are Users Too'', and ITL aims to expand the set of people who can ``program'' machines by replacing coding with natural interactions.
\citet{human_itl} analyze ITL from a human-centered perspective in the context of multi-modal teaching interactions, finding, among other things, that teachers naturally use non-verbal \textit{gestures} alongside verbal instruction.
SUGILITE \citep{sugilite} applies ITL to task automation for smartphone users via the \textit{demonstration} modality. 
The Apprentice Learner \citep{weitekamp2020interaction,maclellan2022domain} acquires task knowledge to power intelligent tutors, learning from both {\it demonstrations} and {\it correctness feedback}. 
In this work, we focus primarily on the \textit{verbal} modality, as we felt recent advances in language-processing transformer models \citep{gpt-4} could significantly augment this arm of interactive task learning.

Past work on ITL using a verbal modality has almost exclusively mapped language to semantics using parsers that are either entirely classical, such as dependency parsers, or utilize a mix of classical and neural methods. We describe some of this past work here, in each case highlighting differences from our approach.

Early work with Instructo-Soar \citep{huffman1993} investigated how to acquire hierarchical task descriptions for use in the Soar cognitive architecture \citep{soar}.
A decade later, this approach has evolved into Rosie \citep{rosie}, a system that can learn game and robotics task knowledge from language instruction.
However, as \citet{rosie2} point out, Rosie's language understanding facility is limited to a ``restricted subset'' of English, limiting user interactions and potentially requiring user training in what Rosie can understand. An example can be seen in Figure~\ref{fig:rosie_chat}, in which Rosie asks for very limited answers to very esoteric questions, e.g., \textit{``What are the verb and parameter arguments associated with this action?''}. More recent continuations of the Rosie project have incorporated LLMs to perform, for example, response generation \citep{kirk2023integrating}. However, this approach still uses a classical parser, rather than incorporating the LLM into its parsing strategy. We seek to remedy this, making interactions with our agent more natural, and requiring less training and mental modeling on the part of the human interlocutor.

PUMICE \citep{pumice} uses a "floating parser" design \citep{floating-parser}, which, while providing a degree of grammatical flexibility, is still bound to exact lexical values and orderings; synonyms, paraphrases, idioms, etc. require manual engineering to overcome. PUMICE, like our system, uses recursive clarification for task learning, and also introduces a demonstration modality, which we have not yet incorporated.

\lstset{language=Lisp,
    basicstyle=\ttfamily\footnotesize,
    stringstyle=\ttfamily\color{black},
    commentstyle=\ttfamily\color{black},
    keywordstyle=\ttfamily\color{black},
    identifierstyle=\texttt,
    tabsize=2
    }
\begin{figure}
\begin{lstlisting}[escapechar=ß,breaklines=true]
U: Cook an onion.
V: How do I cook an onion?
U: First, get an onion. Then, put it in the pot and turn it on.
V: How do I get an onion?
U: Go to the onion and hit space.
...
\end{lstlisting}
\caption{An excerpt of an interaction with VAL to teach the plan shown in Figure~\ref{fig:val_dialog}. \textbf{U} denotes a user message, and \textbf{V} denotes a VAL message.}
\Description{A conversation with VAL. The user and VAL discuss how to cook an onion; the user breaks the task down into steps, which VAL begins to recursively seek clarification for.}
\label{fig:val_chat}
\end{figure}

Building on PUMICE, ONYX \citep{onyx} is a task teaching system situated primarily in a data visualization application. Users teach ONYX to perform data visualization operations in that environment utilizing a mix of demonstration and natural language instruction. The ONYX natural language modality is capable of handling vague or misunderstood instructions by prompting the user to perform a demonstration instead. ONYX uses Microsoft's LUIS language understanding service \citep{luis} to perform intent classification, but in understanding the content of the request, it uses a bigram dependency kernel \citep{bigram-dep} based on a classical dependency parse. ONYX requires a pre-constructed natural language lexicon. While we seek to improve on ONYX's natural language understanding component, we do not yet target the demonstration modality which ONYX uses as a fallback for language misunderstanding.

\citet{suddrey} introduce a system similar to ours, in which hierarchical task structures are acquired and generalized from dialog via recursive clarification. Their system also automatically learns preconditions and effects for actions, which our system does not yet. However, their use of an OpenCCG parser and manually constructed, domain-specific lexicon introduce the same dialog limitations as most other language-based ITL systems.

VAL's niche is linguistic flexibility; while these other systems include features we do wish to add to VAL, like condition learning and additional instructional modalities, we seek here mainly to establish that the careful use of large language models in dialog understanding presents a much more flexible and holistic alternative to the complex and brittle parsing and grounding systems in common use today. Our implementation of VAL also provides, and studies, user-centric features of task learning, such as an undo button and a real-time display of currently known procedures, which have not been explored in prior ITL systems.

\subsection{Differences Between VAL and Prompting-Based Approaches}
\label{sec:related_work}

There are several HCI research efforts actively investigating how non-technical users might leverage LLMs to create AI-powered systems. 
Perhaps most related is work exploring how end users might leverage prompt {\it chaining} (linking multiple LLM prompts together) to create custom chatbot behaviors. For example, prior work has explored the development of tools to support end users in creating prompt chains \citep{wu2022ai,wu2022promptchainer} and investigated users' experiences with using prompting to create chatbots \citep{herdingAICats}. This research is fundamentally different from ITL, but shares the common goal of enabling end users to create and modify the behaviors of AI systems.
Our approach of prompting an LLM to perform specific, narrowly defined functions (e.g., predicate or argument selection) does share some resemblance to the concept of prompt chaining. 
However, there are several key differences between VAL and this prior work.

\begin{figure*}[ht!]
    \begin{center}
    \includegraphics[width=0.7\textwidth]{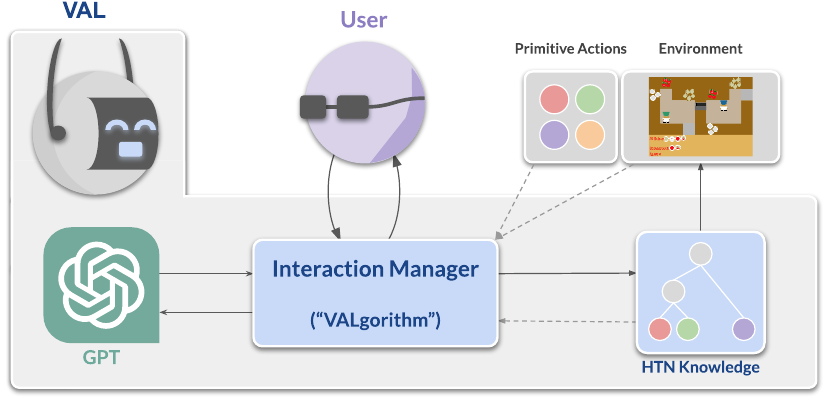}
    \end{center}
    \caption{A high-level diagram of VAL's components: the GPT subroutines (Section~\ref{sec:gpt_subroutine_arch}), the main ``VALgorithm'' (Section~\ref{sec:valgorithm}), the HTN knowledge base (Section~\ref{sec:htns}), and, outside of VAL, an example environment from our user study (Section~\ref{sec:overcooked}).}
    \Description{An architectural diagram of VAL's main components. A central interaction manager interfaces with GPT and with an HTN knowledge base, as well as with the external environment.}
    \label{fig:val_block}
\end{figure*}

First, while VAL has a chat-based interface, it is not a chatbot and it is not controlled with prompting. As highlighted by our need for programmatic scolding (see section~\ref{sec:scolding_results}), prompting provides a poor interface for reliable control of behavior. \citet{zamfirescu2023johnny} found that end users often struggle to create prompts that produce desired behaviors, citing users' low LLM literacy and their tendency to use strategies that resemble human-to-human instruction as a key barrier to overcome. Unlike chatbots, VAL possess the ability to perceive and act directly in the target environment and it has an intentional, task orientation (it learns and executes users' tasks). It also utilizes an interaction design that builds on prior ITL research \citep{rosie, nti} to directly leverage users' prior experience with teaching people. This enables users to engage in productive and natural transfer of task knowledge, even when they have limited LLM and prompting expertise. 
  
Additionally, VAL is more than a chain of LLM prompts. It is a neuro-symbolic hybrid AI {\it system} (see section~\ref{sec:gpt-integration-philosophy}), consisting of a fully-specified knowledge representation (HTNs), performance component (which grounds HTN tasks into action), and learning component (see Algorithm~\ref{alg:valgorithm}). This enables it to overcome several limitations highlighted in prior work. For example, \citet{wu2022ai} highlight that LLMs lack key reasoning capabilities and \citet{herdingAICats} point out that it is difficult to get LLMs to reliably say ``I don't know''. In contrast, VAL can leverage its symbolic HTN knowledge to engage in reliable multi-step reasoning, and it can recognize when it does not know how to do something, so it can ask for additional guidance.

Another major difference between VAL and prompt-based LLM approaches is that  while prompting does enable user customization (based on the contents of the prompt), it does not enable the system to actually learn from the user.\footnote{Fine tuning can extend an LLM's knowledge, but comes with its own set of challenges and is usually not employed by end users.} The training set for most LLMs is enormous, but many tasks are idiosyncratic and will not be well represented in the training data. For example, when we tested our LLM model directly on Overcooked tasks, we found it would make up irrelevant, but plausible actions, such as washing the knife (see section~\ref{sec:gpt-integration-philosophy}). This is because the LLM possesses general cooking knowledge (which apparently emphasizes washing utensils), but not the idiosyncratic knowledge of how to cook in the Overcooked environment. In contrast, VAL utilizes ITL to learn novel task knowledge on the fly from its interactions with users, even if these tasks are not well represented in the LLM model's training data. This enables VAL to be successful in situations where prompt-based approaches would fail.




\section{VAL System Design}
\label{sec:design}

VAL acquires hierarchical task knowledge by exploiting the hierarchical nature of clarification in dialog. Users give VAL commands, and when VAL is unable to relate that command to existing task knowledge, it asks the user to define the steps of the command (see Figure~\ref{fig:val_chat}). This process continues recursively until all tasks have been ``grounded out'' in terms of known actions, initially comprising a set of environment-dependent primitive actions.

To facilitate natural language as a medium for these dialogs, VAL uses an array of GPT-powered subroutines to perform tasks like action segmentation, predicate selection, and argument grounding. However, these subroutines are largely \textit{translational} between the user input and the underlying symbolic knowledge structures; the overall task learning algorithm itself is not executed by GPT, and when VAL acts in the environment, or when users want to audit VAL's knowledge, the black-box GPT subroutines are not involved.

While VAL's recursive task acquisition mechanism is inspired by prior work, this synthesis of GPT with the symbolic task acquisition process is its main contribution, and hopefully serves as another step toward truly general neuro-symbolic hybrid systems.

\begin{figure}
    \includegraphics[width=\columnwidth]{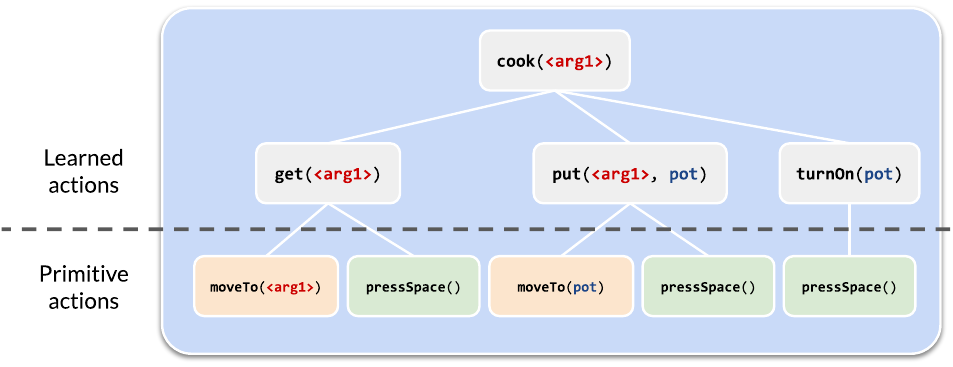}
    \caption{An example HTN plan for cooking, learned by VAL in the Overcooked-AI environment. The learned task \protect\texttt{cook} decomposes into other learned tasks, which themselves decompose into primitive actions.}
    \Description{An example HTN plan for cooking, learned by VAL. The cook action breaks down into get, put, and turnOn actions, each of which break down further into the atomic commands moveTo and pressSpace.}
    \label{fig:htn}
\end{figure}

\subsection{Knowledge Representation}
\label{sec:htns}

VAL's task knowledge representation is based on the Hierarchical Task Network (HTN) formalism \citep{htns}, a tree structure in which tasks decompose into sequences of subtasks, each task contains preconditions and effects, and leaves are ``primitives''  given as part of the environment. HTNs are widely applicable to planning, and have been used in many domains, from video games \citep{htn-game} to user preference modeling \citep{htn-userpref} to transport logistics \citep{htn-shop2}.

An example of a VAL HTN in the cooking video game domain of our user study is shown in Figure~\ref{fig:htn}. While the general HTN formalism includes precondition and effect modeling, this study only focuses on evaluating the acquisition of non-conditional hierarchical task structures from dialog.


\subsection{Hierarchical Task Acquisition}

Tasks are conveyed from the user to VAL starting with an initial command provided by the user. The command can be an action or a sequence of multiple actions. All actions represented in the command are matched to corresponding predicate-argument action structures already learned by VAL, including the initial primitive actions. If any actions in a command cannot be matched to semantically similar known actions, they are recursively clarified with new subcommands solicited from the user; the recursive clarification of actions into sequences of new actions forms the hierarchical structure of the learned task. One step of this process is depicted in Figure~\ref{fig:teaser}, and the full procedure is described in Algorithm~\ref{alg:valgorithm}.

Once tasks are acquired, they are added to VAL's knowledge base and can be chosen and executed for future requests. These requests fill in their generalized values and traverse the task tree to the leaves to send primitive actions to the environment. For more information on this process, see Section~\ref{sec:realtime}.

\subsection{Plan Generalization}

When new plan knowledge is acquired, a GPT subroutine is used to determine which arguments to generalize and which to hold constant. This means that learned tasks can be reused, e.g., after being told how to cook an onion in the Overcooked video game environment (see Section~\ref{sec:overcooked}), VAL should be able to cook a tomato without seeking further clarification. The generalization component, \path{genGPT}, is described in Section~\ref{sec:gpt_subroutine_arch}.

\subsection{GPT Integration Philosophy}
\label{sec:gpt-integration-philosophy}

VAL is an AI \textit{system}, rather than just a \textit{model}: it comprises a combination of GPT components, a classical learning algorithm, and connecting code to interface with a given environment. We use GPT models to solve atomic, linguistic subroutines within the well-defined ``VALgorithm''. This allows us to obtain more focused, reliable output than if we had used language models to solve the entire task end-to-end. It also allows us to perform better error correction on a more predictable set of error classes. When language models alone are used for this task, without the entire VAL architecture, they frequently invent new actions not yet learned, or attempt to fabricate non-applicable domain knowledge, such as washing a knife in the Overcooked video game despite the game lacking any washing operations.

The tasks performed by the language model here, though, are by no means trivial; extraction of predicate-argument structures from text, and the mapping of those structures to known actions, both require resolution of a considerable number of linguistic tasks, such as syntactic parsing, anaphora resolution, paraphrase resolution, action segmentation, and more; the full set of tasks performed by GPT in VAL is detailed in Section~\ref{sec:gpt_subroutine_arch}. 

\begin{figure}

\makeatletter
\newcommand{\removelatexerror}{\let\@latex@error\@gobble}
\makeatother

\begingroup
\removelatexerror
\begin{algorithm}[H]
    \label{alg:valgorithm}
    \DontPrintSemicolon
    \SetKwFunction{interpret}{interpret}
    \SetKwFunction{segmentGPT}{segmentGPT}
    \SetKwFunction{mapGPT}{mapGPT}
    \SetKwFunction{newAction}{newAction}
    \SetKwFunction{groundGPT}{groundGPT}
    \SetKwFunction{verbalizeGPT}{verbalizeGPT}
    \SetKwFunction{paraphraseGPT}{paraphraseGPT}
    \SetKwFunction{nameGPT}{nameGPT}
    \SetKwFunction{genGPT}{genGPT}
    
    \SetKwProg{Fn}{Function}{:}{}
    
    \Fn{\interpret{\beigegptbox{input}  , \redgptbox{defs} }}{
        steps $\leftarrow$ [ ]\;
        \For{\greengptbox{seg} \textbf{in} \scalebox{0.85}{ \gptbox{segmentGPT} (\beigegptbox{input} )} }{
            \greengptbox{pred} $\leftarrow$ \scalebox{0.85}{ 
            \gptbox{mapGPT} (\beigegptbox{seg} , \redgptbox{defs} ) } \;
            \If{\greengptbox{pred} is None}{
                steps $\leftarrow$ steps + \newAction(\beigegptbox{seg} , \redgptbox{defs} )\;
            }
            \Else{
                \greengptbox{args} $\leftarrow$ \scalebox{0.85}{ \gptbox{groundGPT} (\beigegptbox{seg} , \greengptbox{pred} , \redgptbox{defs} ) } \;
                \greengptbox{v} $\leftarrow$ \scalebox{0.85}{ \gptbox{verbalizeGPT} (\greengptbox{pred} , \greengptbox{args} ) } \;
                \If{ \scalebox{0.85}{ \gptbox{paraphraseGPT} (\greengptbox{v} , \beigegptbox{seg} ) } }{
                    steps $\leftarrow$ steps + ( \greengptbox{pred} , \greengptbox{args} )\;
                }
                \Else{
                    steps $\leftarrow$ steps + \scalebox{0.85}{\newAction(\beigegptbox{seg} , \redgptbox{defs} )}\;
                }
            }
        }
        \Return{steps}\;
    }
    \BlankLine
    \Fn{\newAction{\beigegptbox{input} , \redgptbox{defs} }}{
        \greengptbox{newPred} $\leftarrow$ \scalebox{0.85}{ \gptbox{nameGPT} (\beigebox{input} ) } \;
        \beigegptbox{expl} $\leftarrow$ askUser(``\textit{How do I }\beigegptbox{input} \textit{?}'')\;
        \greengptbox{defn} $\leftarrow$ \interpret(\beigegptbox{expl} )\;
        \greengptbox{newArgs} $\leftarrow$ \scalebox{0.85}{ \gptbox{genGPT} (\greenbox{newPred} , \greenbox{defn} ) } \;
        \redgptbox{defs} [\greengptbox{newPred} ] $\leftarrow$ (\greengptbox{newArgs} , \greengptbox{defn} )\;
        \Return{(\greengptbox{newPred} , \greengptbox{newArgs} )}\;
    }

    \BlankLine
    \vspace{10pt} \caption{The ``VALgorithm'' governing VAL's high-level task acquisition mechanism. All use of GPT is abstracted into narrowly-scoped subroutines.}
    \Description{Pseudocode for the ``VALgorith'' managing VAL's interaction state. GPT subroutine calls are illustrated in rounded purple boxes.}
\end{algorithm}
\endgroup

\end{figure}

Our integration of GPT into an otherwise classical algorithm has allowed us to capitalize on the \textit{fluency} of large language models without falling victim to the myriad \textit{failures} that can arise from using them to complete non-linguistic tasks end-to-end---especially recursive tasks and tasks requiring well-defined output.

\subsubsection{Programmatic Scolding}
\label{sec:scolding}
As we developed VAL, we moved between two language model bases for the GPT components: \path{gpt-3.5-turbo} and \texttt{gpt-4}. These models frequently exhibit overly cautious behavior: for example, \path{gpt-3.5-turbo} will often refuse to choose cooking-related actions for cooking-related inputs in the Overcooked game setting, citing its lack of physical body as an AI language model as a reason for its inability to cook food.

This was quite frustrating, and given the black-box nature of the models and the great influence of reinforcement learning from human feedback on their training \citep{rlhf}, stymied us for some time. Eventually, we settled on a silly but effective solution we deem \textit{programmatic scolding}. In programmatic scolding, responses containing apology words or the string \textit{as an AI language model} are automatically detected and responded to with a message scolding the language model for being culturally insensitive in its refusal to answer. We found these additional scolding prompts would lead the language model to produce the desired output. Literally scolding the model to produce well defined outputs is not a desirable research strategy, but underscores the difficulty of controlling LLM outputs.


\section{VAL System Architecture}
\label{sec:system}

\subsection{The VALgorithm}
\label{sec:valgorithm}

Like in the hierarchical task acquisition work by \citet{suddrey}, VAL acquires tasks from natural language descriptions by grounding actions and arguments and recursively seeking new definitions for any unknown actions in terms of known ones. However, aside from the GPT subroutines that make VAL's implementation of these actions unique, the actual VAL algorithm (``VALgorithm'') differs in its own ways, and so we describe it here, and in the pseudocode of Algorithm~\ref{alg:valgorithm}.

VAL begins by segmenting the input into individual, temporally ordered steps. Each of these steps is then mapped independently to known actions. If a suitable match is found, arguments are selected to fill its slots. The filled predicate-argument representation is then verbalized into a sentence, and that sentence is compared with the initial input segment; if they are determined to be paraphrases of one another, the mapped action and filled argument slots are used. However, if the matched predicate is deemed too semantically different from the input segment after matching, or if no match is initially found, then a definition is sought for it, a name is created for the new action, and a subset of the used arguments are selected as the variable arguments for the new action.

Each \gptbox{purple symbol} ~ in Algorithm~\ref{alg:valgorithm} represents a GPT subroutine, each of which is defined below, in Section~\ref{sec:gpt_subroutine_arch}.

\subsection{GPT Subroutines}
\label{sec:gpt_subroutine_arch}

While the VALgorithm is straightforward and interpretable on its face, its subroutines encapsulate much of the complexity inherent to any natural language understanding task. Below, we enumerate and describe each of the GPT subroutines essential to VAL's ability to understand naturally phrased instructions and relate them to its symbolic models of the environment and of actions within it. We implemented each of these subroutines as a prompt, occasionally with a set of examples; we did not use fine-tuning for any GPT subroutines.

\newpage


\subsubsection{\textbf{\protect\path{segmentGPT}}}\

\vspace{0.3cm}
\includegraphics[width=0.96\columnwidth]{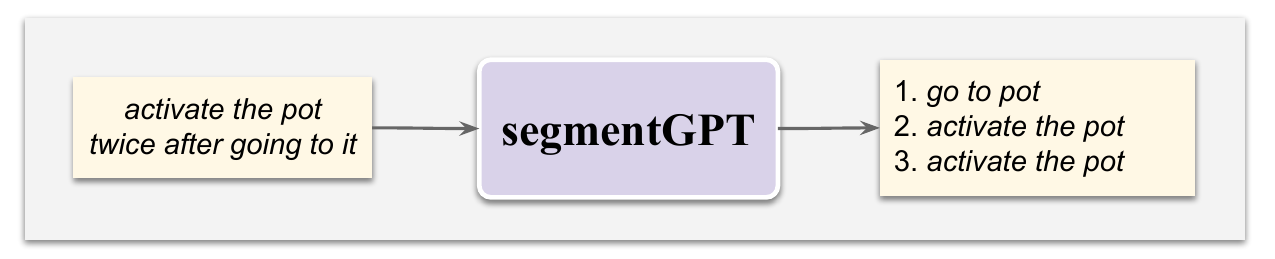}
\vspace{0.01cm}

\path{segmentGPT} splits a command into atomic steps. Because each atomic step will be matched individually to known actions, co-references must be resolved as well, so that pronouns in later steps are replaced with their referents. \path{segmentGPT} also expands repeated actions, such as those with adverbs like ``twice'', into their own steps, and orders the steps chronologically, resolving temporal relations like ``before'' and ``after'' in the input.

\vspace{0.1cm}

\subsubsection{\textbf{\protect\path{mapGPT}}}\

\vspace{0.3cm}
\includegraphics[width=0.96\columnwidth]{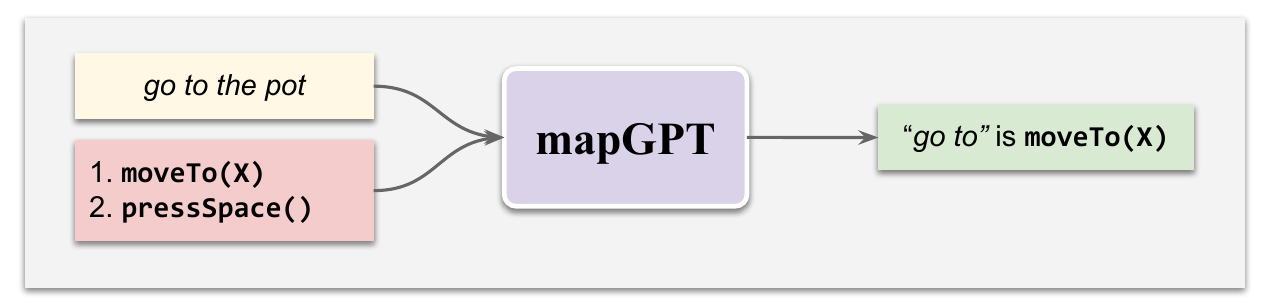}
\vspace{0.01cm}

\path{mapGPT} selects a known action corresponding to an input segment produced by segmentGPT, or determines that no known action is suitable. In the latter case, VAL will attempt to acquire a new action (or, if confirmatory dialogs are enabled, ask if a suitable matching action exists). \path{mapGPT} does not jointly select arguments for the chosen action; that subsequent task is factored out and performed by \path{groundGPT}.

\vspace{0.1cm}

\subsubsection{\textbf{\protect\path{groundGPT}}}\

\vspace{0.3cm}
\includegraphics[width=0.96\columnwidth]{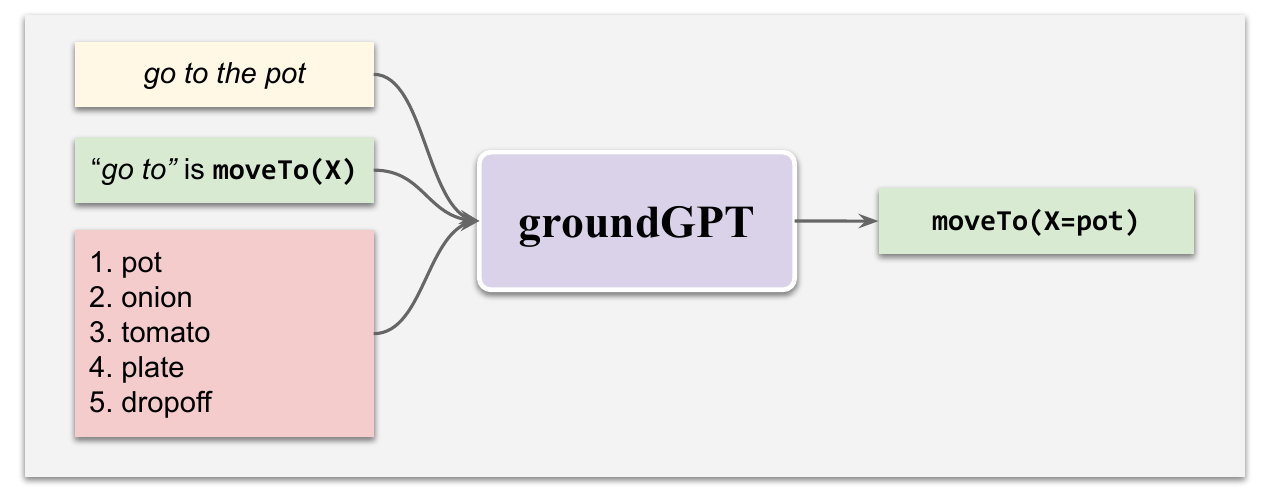}
\vspace{0.01cm}

\path{groundGPT} selects from a list of known world objects to fill argument slots for a predicate chosen by \path{mapGPT}. \path{groundGPT} also has access to the original text segment. \path{groundGPT} must correctly interpret plurals, restrictors, and other quantifiers in the input: for example, a single noun phrase like ``the numerators'' may expand to fill two argument slots, or ``any red vegetable'' may be used to select ``tomato''. \path{groundGPT} only selects arguments when a known action has been chosen first; to choose a set of arguments for a newly introduced action, VAL uses \path{genGPT}.

\newpage

\subsubsection{\textbf{\protect\path{verbalizeGPT}} and \textbf{\protect\path{paraphraseGPT}}}\

\vspace{0.3cm}

    \includegraphics[width=0.96\columnwidth]{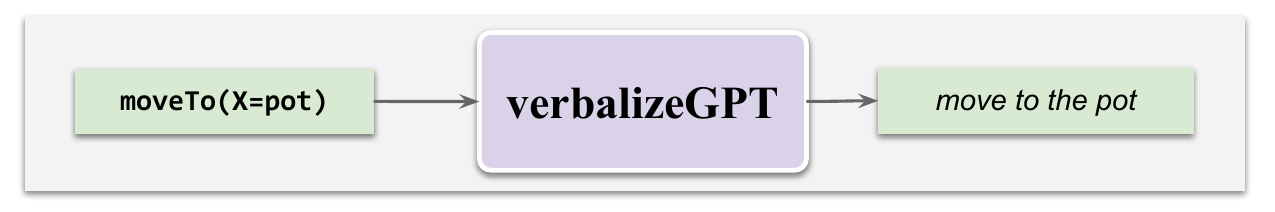}

    \vspace{5mm}
    \includegraphics[width=0.96\columnwidth]{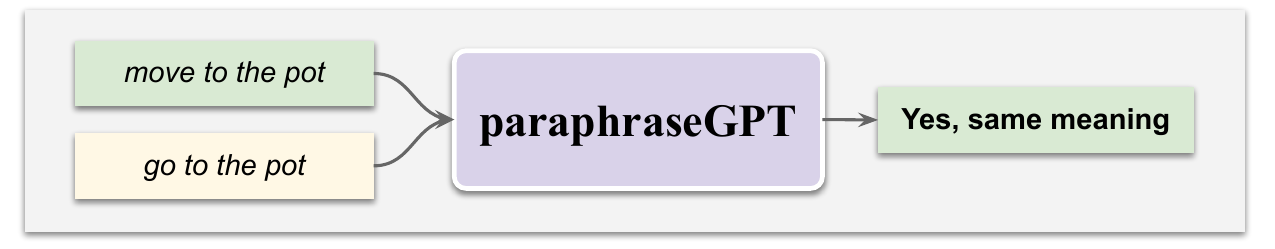}

    \vspace{5mm}

\path{verbalizeGPT} is used in tandem with \path{paraphraseGPT} to verify matches before committing to them; occasionally, \path{mapGPT} will select an action with an inexact semantic match, such as the selection of the action \path{moveTo} for the input segment ``get an onion''. The reasons for this are not totally clear\footnote{However, when we extend the \texttt{mapGPT} prompt to a conversational setting with ChatGPT and ask why it chose what it did, it will often justify it by saying things like ``in order to get an onion, you must first go to it''. It is unclear, though, whether this reasoning informed its initial choice or is merely a consequence of needing to justify it retrospectively.}, but we find that these incorrect choices can often be avoided by the following three-step process:

\begin{enumerate}
    \item Verbalize the action and arguments chosen by \path{mapGPT} and \path{groundGPT} into a natural sentence using \path{verbalizeGPT}.
    \item Compare that sentence to the original user input segment using \path{paraphraseGPT}.
    \item Accept the match if and only if \path{paraphraseGPT} deems them a match.
\end{enumerate}

When confirmatory dialogs are enabled, \path{verbalizeGPT} and \path{paraphraseGPT} are not used, as the user provides suitability information directly.

\vspace{0.1cm}

\subsubsection{\textbf{\protect\path{nameGPT}}}\

\vspace{0.3cm}

\includegraphics[width=0.96\columnwidth]{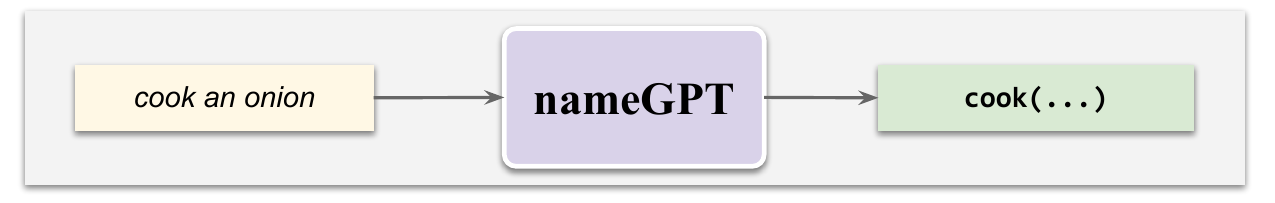}

\vspace{0.1cm}

\path{nameGPT} produces a predicate name for the original text snippet responsible that led to the currently unnamed task being learned.
If \path{nameGPT} produces a name identical to a known name, progressively higher numbers are appended, beginning with 2; however, this is a shallow solution to a relatively deep problem, as we will discuss in Section~\ref{sec:discussion}.

\vspace{0.1cm}

\subsubsection{\textbf{\protect\path{genGPT}}}\

\vspace{0.3cm}

\includegraphics[width=0.96\columnwidth]{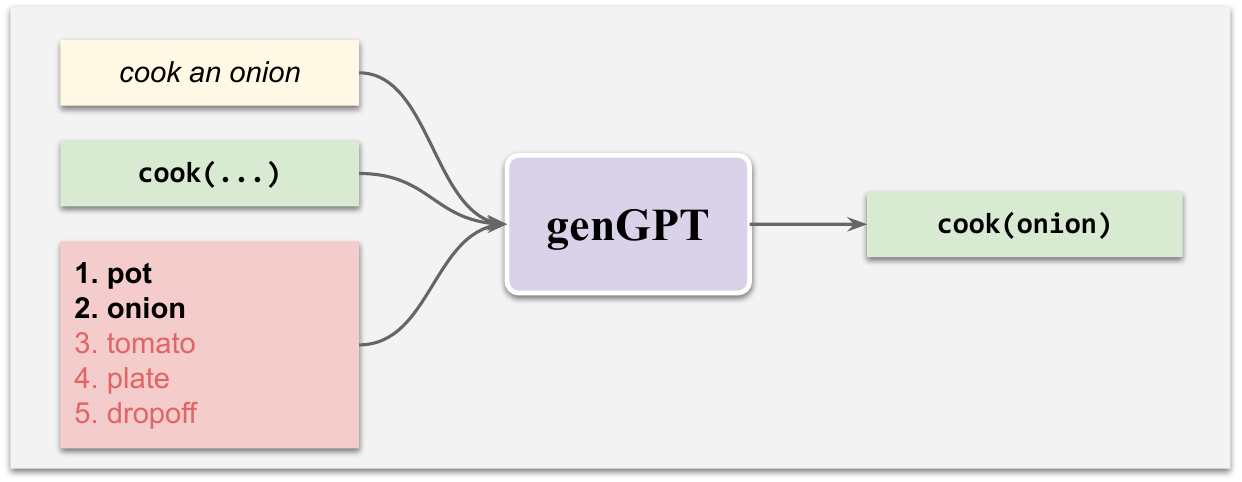}

\vspace{0.1cm}

\path{genGPT} selects arguments for newly learned actions after \path{nameGPT} decides on a predicate name for them. It is only allowed to choose from the set of arguments that were \textit{used} in the full expansion of the task. We use GPT to filter this set because selecting which arguments should be generalized and which should remain constant is a task that requires both lexical and semantic knowledge. However, \path{genGPT} occasionally omits an important argument, leaving it constant; or includes an argument that will never vary, such as the pot in the \texttt{cook} action.

\section{E``VAL''uation}
\label{sec:study}

\begin{figure}
    \centering
    \includegraphics[width=\columnwidth]{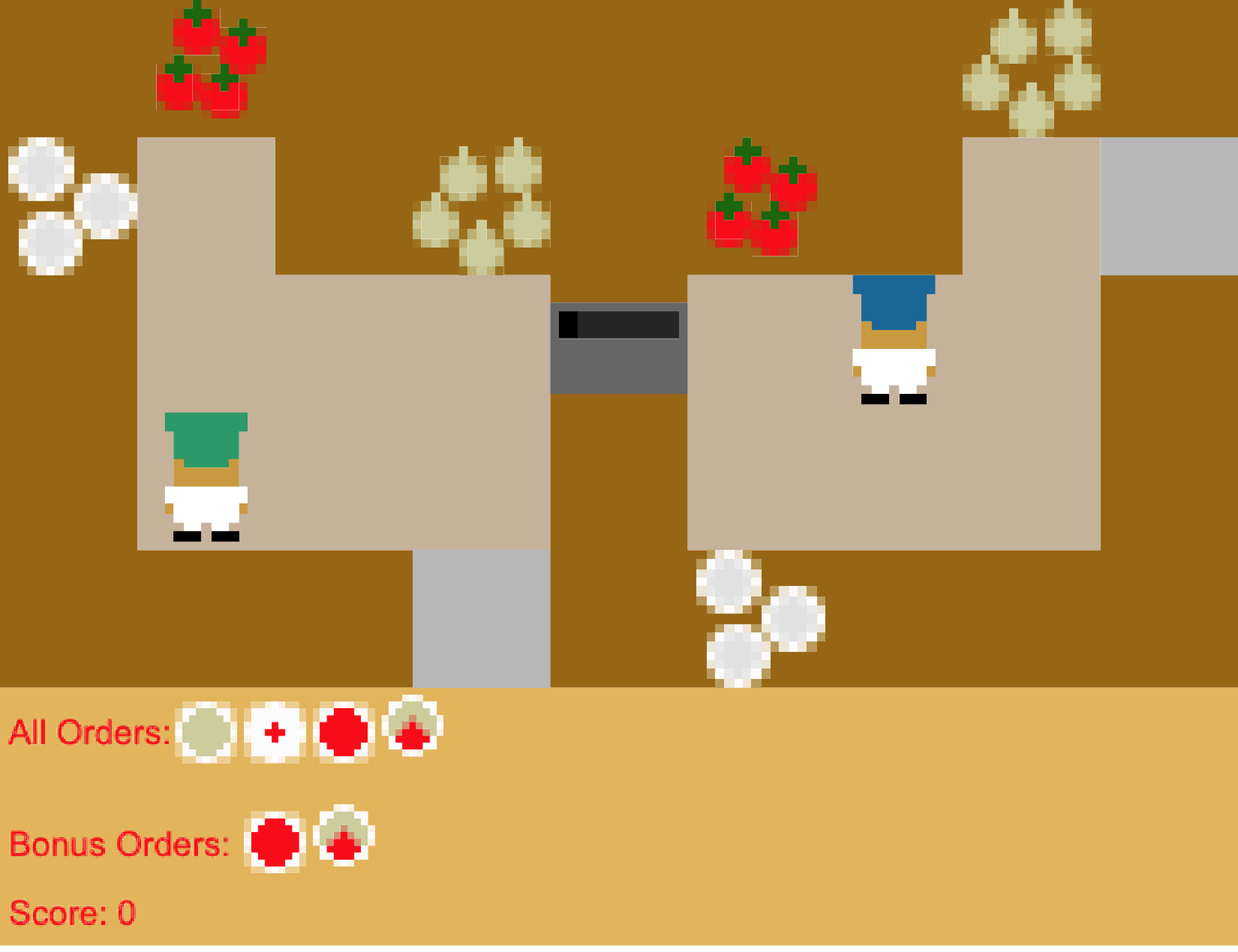}
    \caption{A screenshot of the Overcooked-AI game environment used in VAL's user studies.}
    \Description{A screenshot of the Overcooked-AI game environment used in VAL's user studies. Two cartoon sprites of chefs sit in kitchen-like environments, with vegetables and cooking tools available to them.}
    \label{fig:oc_game}
\end{figure}

To evaluate VAL, we conducted an IRB-approved user study, in which we recruited 12 participants to teach VAL a task in a video game environment and take a survey regarding their experience.\footnote{We chose this number of participants as it was found to be the most common participant sample size in CHI submissions as of 2016. \citep{chi_12_participants}} This study had two aims:
\begin{enumerate*}
    \item to evaluate the participants' subjective experiences via the survey, and
    \item to evaluate the objective performance of VAL's components via signals recorded both from the interaction logs and from VAL's implementation.
\end{enumerate*}

In the latter case, we largely relied on \textit{confirmatory dialogs} (see Section~\ref{sec:confirm}), via which users either approve or correct decisions made by VAL's components, as a source of ground truth for measuring component success. We also measured signals for in-game actions successfully completed by VAL, instances of users clicking ``undo'', and any system crashes. We report and analyze the survey results in Section~\ref{sec:survey_results}, and all other results in Section~\ref{sec:metric_results}.\footnote{VAL's code and prompts can be accessed at \url{https://github.com/bitbanger/overcooked-demo/tree/paper_mess}.}

\subsection{Environment}
\label{sec:overcooked}

For our user study environment, we chose Overcooked-AI \citep{overcooked-ai}, an open-source, simplified implementation of the popular video game Overcooked \citep{overcooked}.
An increasing amount of contemporary task learning and task performance work has focused on this environment \citep{proagent,coach,overcooked-gen}.
After showing each user a tutorial for the game and allowing them to practice the task manually, we asked them to teach VAL to make onion soup with one onion, which involves putting the onion in the pot, turning the pot on, fetching a plate, interacting with the pot while holding the plate, and taking the plate to the drop-off location.

\subsubsection{Primitive Actions}
As initial primitive actions, we implemented \texttt{moveTo} and \texttt{pressSpace}. While the latter straightforwardly lives up to its name, we chose to implement the former as a breadth-first search within the environment; while work has been done on HTN representations of pathfinding \citep{htnpath}, we decided that verbal acquisition of such networks would be too complex for average users to teach in an initial study. The selection of primitives is an important question in integrating VAL with a new environment; it is important to find a balance between ease of implementation and ease of teaching. In this case, we felt that the task of making soup remained suitably interesting with low-level movement primitives introduced.

\begin{figure}
    \centering
    \includegraphics[width=0.96\columnwidth]{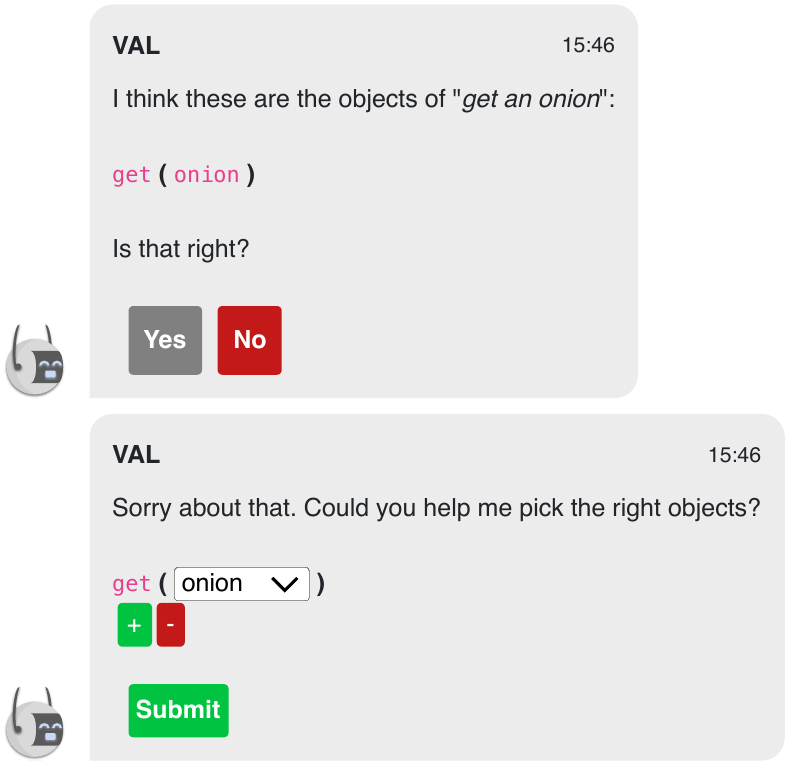}
    \caption{A confirmatory dialog example for \protect\gptbox{genGPT} . Examples of all types of confirmatory dialog are available in Appendix~\ref{app:confirm_dialogs}.}
    \Description{An example confirmatory dialog used in the VAL user study. VAL asks questions to confirm its GPT subroutine results, and the user responds ``yes'' or ``no'', optionally providing a manual correction in the event they select ``no''.}
    \label{fig:confirm_dialogs}
\end{figure}

\subsection{Study Design Features}

\subsubsection{Confirmatory Dialogs}
\label{sec:confirm}
Interactive task learning is a sequential process in which an early error can compromise all downstream steps and the final result. Furthermore, we required a way to evaluate the performance of VAL's individual GPT subroutines relative to human judgment. To address both of these issues, the VAL user study makes use of confirmatory dialogs for GPT subroutine results. These prompt the user to either confirm or correct the result returned by the subroutine. Examples of confirmatory dialogs can be seen in Figure~\ref{fig:confirm_dialogs} and Appendix~\ref{app:confirm_dialogs}.

As these dialogs interrupt the flow of the interaction and may frustrate users with their frequency and repetitiveness, it would be ideal not to ever require them in a non-study setting. In Section~\ref{sec:gpt-rates}, we discuss a strategy that seems to reduce the need for these dialogs in one component, and our plans to pursue similar strategies for other components.

The VAL dialog interface was adapted from the Simple Chat UI system developed by Sajad Hashemian \citep{simplechat}. We contributed the design and implementation of the confirmatory dialogs within the Simple Chat UI message bubbles, which contain only text by default.

\subsubsection{Knowledge Display}
\label{sec:knowledge_display}

\begin{figure}
    \centering
    \includegraphics[width=0.5\columnwidth]{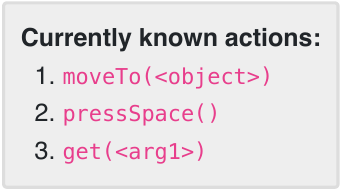}
    \caption{The knowledge display (see Section~\ref{sec:knowledge_display}) shown on the study page. In this example, \protect\texttt{moveTo} and \protect\texttt{pressSpace} are primitive actions, and \protect\texttt{get} is a learned action taught by the user.}
    \Description{An enumeration of three currently known actions from a VAL interaction: moveTo, pressSpace, and get.}
    \label{fig:knowledge_display}
\end{figure}

At all times, a knowledge display box (see Figure~\ref{fig:knowledge_display}) was visible on the study page, showing which generalized actions had been learned by VAL, including both primitive and learned actions.

\subsubsection{Real-Time Action Performance}
\label{sec:realtime}
As VAL grounds the task out in terms of primitive actions, a decision must be made as to when to send each primitive action to the game environment.
In early testing, users found it confusing to wait until VAL had learned an entire complex task tree before it took any actions in the game.
Therefore, we opt to send each primitive as it's added to the tree, rather than sending all primitives once the tree has been completed.
This allows the user to monitor the task progress during instruction and confirm that the right actions are being carried out.

\subsubsection{The Undo Button}
To further enable error correction during user sessions, we implemented an undo button in the VAL dialog interface, which is visible in Figure~\ref{fig:val_dialog}.
David R. Hill \citep{undo} offered the guidance that, in a dialog setting with external effects, what should be undoable is each \textit{effect}, rather than each \textit{message}.\footnote{Prof. Hill also noted that an undo feature ``takes a lot of effort to implement'', which finding is corroborated by our work.}
In a VAL session, all modification of game or agent knowledge state is a reaction to a user input, whether that be a command or a confirmatory dialog response.
Therefore, when the undo button is clicked, we opt to rewind both the game and agent to their states at the previous prompt for user input.

\subsection{Experimental Setup}
We recruited 12 participants by word of mouth from university communities.
After signing consent forms, they were given two static webpage tutorials: one for making soup in the Overcooked-AI environment, and one for teaching an unrelated task to VAL.
In between the two tutorials, they were also asked to practice making soup manually in the game environment.
After completing both tutorials, the participants were asked to instruct VAL to make onion soup, and then presented with a side-by-side view of a VAL dialog (e.g., Figure~\ref{fig:val_dialog}), a display of VAL's knowledge (see Figure~\ref{fig:knowledge_display}), and the game (e.g., Figure~\ref{fig:oc_game}).
The participants were informed that they could end the study at any time, regardless of progress, and proceed to the survey.

The post-session survey asked non-identifiable questions about relevant prior experience, seven-point Likert scale opinion questions, and open-ended short-answer opinion questions. The full survey form is attached as Appendix~\ref{app:survey}, and a link to the full, anonymized set of results, including short-answer responses, will be provided if and when this paper is no longer under blind review.

\subsection{User Survey Results}
\label{sec:survey_results}
\begin{figure*}
    \includegraphics[width=\textwidth]{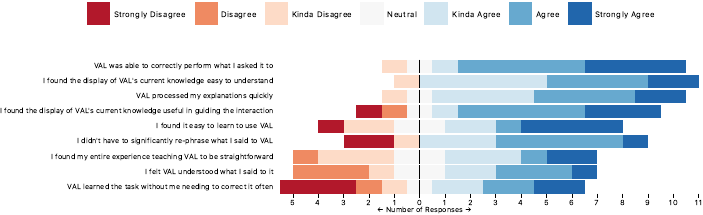}
    \caption{Participant Likert scale scores for the questions asked in our post-study survey. Red/left is bad, and blue/right is good. Tick marks are sometimes aligned at half-intervals due to neutral responses straddling the center line.}
    \Description{A representation of the user-provided Likert scores for usability questions in the VAL user study. The figure shows that the majority of responses indicated positive sentiments regarding VAL's usability.}
    \label{fig:val_q4}
\end{figure*}

\subsubsection{Prior Experience}

We asked each participant about their prior experience levels with interacting with large language models, computer programming, and video games like Overcooked-AI. All participants had at least ``a lot'' of experience with programming; most participants only had ``a little'' experience interacting with large language models; and most participants had at least ``a moderate amount'' of experience with video games like Overcooked-AI. The full results for these questions are presented in Appendix~\ref{app:prior_exp_results}.

\subsubsection{Opinions on Sessions}

The questions we asked participants about their sessions were derived in part from prior work on task learning system usability, although some were original to this work.
While some questions are inspired by prior work, the questions are highly subjective and situational, and so responses cannot be directly compared as if they were on the same scale; what is expected or acceptable for a user in one context may not be in another, and questions about, e.g., overall ease of use, mask too much context for direct numerical response comparison.

The prompts we gave, and any applicable works inspiring those questions, were:

\begin{enumerate}
    \item \textit{VAL was able to correctly perform what I asked it to} \citep{nti-woz}
    \item \textit{I found the display of VAL's current knowledge easy to understand} \citep{onyx}
    \item \textit{VAL processed my explanations quickly} \citep{nti-woz}
    \item \textit{I found the display of VAL's current knowledge useful in guiding the interaction} \citep{onyx}
    \item \textit{I found it easy to learn to use VAL} \citep{nti-woz, onyx}
    \item \textit{I found my entire experience teaching VAL to be straightforward}
    \item \textit{I felt VAL understood what I said to it}
    \item \textit{VAL learned the task without me needing to correct it often}
\end{enumerate}

The responses are visualized in Figure~\ref{fig:val_q4}.
For the most part, user experience was positive; all questions had more agree-type responses than disagree-type responses.
In keeping with the observed game milestone signals, almost all participants reported being able to complete the task.
The worst responses were to the prompts \textit{VAL learned the task without me needing to correct it often} and \textit{I felt VAL understood what I said to it}; this highlights the need for the confirmatory dialogs found in the study, although finding ways to reduce them may still be beneficial.
In fact, the high number of confirmatory dialogs may have contributed to the slightly higher proportion of negative scores for \textit{I found my entire experience teaching VAL to be straightforward} as well.

\subsection{System Metric Results}
\label{sec:metric_results}
\setlength{\fboxsep}{4pt}
\setlength{\fboxrule}{0pt}
\begin{table}[ht]
\resizebox{0.96\columnwidth}{!}{
\begin{tabular}{ c | c }
  \fbox{\textbf{GPT Subroutine}} & \fbox{\textbf{Success Rate}} \\
  \Xhline{2\arrayrulewidth}
  \gptbox{segmentGPT} & \makecell[l]{\textbf{93\%} user approval} \\ 
  \hline
  \gptbox{mapGPT} & \makecell[l]{\textbf{82\%} user approval (\texttt{gpt-3.5-turbo}) \\ \textbf{97\%} user approval (\texttt{gpt-4})} \\ 
  \hline
  \gptbox{groundGPT} & \makecell[l]{\textbf{88\%} user approval} \\ 
  \hline
  \gptbox{genGPT} & \makecell[l]{\textbf{81\%} user approval} \\ 
  \hline
  \gptbox{verbalizeGPT} \hspace{0pt} + \gptbox{paraphraseGPT} & \makecell[l]{\textbf{79\%} true positive rate \\ \textbf{99\%} true negative rate} \\
\end{tabular}
}
\caption{GPT subroutine success rates across all the times they were activated within the twelve user sessions.}
\Description{A table displaying success rates for the GPT subroutines used by VAL.}
\label{tab:gpt-rates}
\end{table}

\subsubsection{GPT Subroutine Success Rates}
\label{sec:gpt-rates}

Table~\ref{tab:gpt-rates} shows the success rates for four of the seven GPT subroutines described in Section~\ref{sec:gpt_subroutine_arch}. We did not measure the success of \texttt{nameGPT}, as predicate name choice was not subject to user approval or disapproval in our study. However, there is clear evidence from question (2) on the opinion survey that users found the display of VAL's current knowledge easy to understand. As the names of the learned tasks in the display were the output of nameGPT, this suggests its outputs are interpretable and in alignment with user's expectations.

The success rates of \texttt{paraphraseGPT} and \texttt{verbalizeGPT} are not independently defined, as these two components work together to filter spurious matches produced by \texttt{mapGPT}. As the design of the study required manual user approval for \texttt{mapGPT} decisions, these two components were unused. However, we calculated the decisions that \textit{would} have been made according to these two components, and compared them to the user decisions that were actually made. We found a false negative rate of 21\% and a false positive rate of 1\% relative to the user-provided ground truth. The low false positive rate suggests that roughly four fifths of confirmatory dialogs for \texttt{mapGPT} outputs may be avoidable, with minimal impact to performance, by simply always accepting negative paraphrase decisions made by these two components.

\subsubsection{Programmatic Scolding Occurrences}
\label{sec:scolding_results}
Programmatic scolding (see Section~\ref{sec:scolding}) was triggered a total of 29 times, split across three participants. All 29 scolds occurred in the \texttt{gpt-3.5-turbo} participant group.

\subsubsection{Game Milestone Signals}
During the sessions, we recorded signals from the game engine corresponding to certain ``milestone'' events: picking up an onion, putting an onion in the pot, turning the pot on, plating the cooked soup, and delivering the soup.

Of 12 total participants, nine were able to instruct VAL to complete every milestone; two were able to instruct VAL to cook the soup, but not to plate or deliver it; and one participant was unable to complete any milestones.

\subsubsection{Undos}
10 of 12 participants used the undo feature. The median number of undos was 6, and the mean was 9.

\subsubsection{System Crashes}
Of the three participants who did not signal all five milestones, two experienced crashes in the VAL system that halted all progress. One participant experienced a crash after completing all milestones.

The root cause of two of these three crashes was a malformed string returned by a GPT subroutine. The third crash was due to a platform-specific string encoding issue in the experimental environment that was not observed in the development environment.

\subsubsection{Choice of Base GPT Model}
\label{sec:gpt-size-perf}

Any autoregressive language model can be used to implement VAL's GPT-based subroutines. Of our 12 participants, seven interacted with a \texttt{gpt-3.5-turbo}-based VAL system, and the other five interacted with a \texttt{gpt-4}-based one.

The switch to GPT-4 had a statistically significant ($p \le .05$) effect\footnote{p-values were calculated using $t$-tests for all metrics except approve-reject decisions, for which we used binary $\chi^{2}$ tests.} on:

\begin{enumerate}
\item the \textbf{frequency of programmatic scolding} (decreased; good)\footnote{In fact, there was no scolding at all in the \texttt{gpt-4}-based sessions.}
\item the \textbf{number of crashes} (decreased; good),
\item the \textbf{success rate of \texttt{mapGPT}} (increased; good),
\item the \textbf{user-perceived processing speed} (increased; bad),
\item the \textbf{ease of learning to use VAL} (increased; good), and
\item the \textbf{straightforwardness of the entire experience} (increased; good).
\end{enumerate}

The other categories tested for significance were:

\begin{enumerate}
\item the number of messages sent by the user per completed milestone,
\item the false positive and false negative rates of \texttt{paraphraseGPT} filtering when compared user decisions,
\item the number of clicks of the ``undo'' button,
\item all GPT subroutines other than \texttt{mapGPT}, and
\item all of the other questions in the survey.
\end{enumerate}

\subsection{Discussion}
\label{sec:discussion}

This section reflects on the quantitative results above and on the short-answer responses given in the survey. We break the discussion into themes digested from these results.

\subsubsection{Naturalness of Interactions}
Most participants responded that they felt empowered to use natural phrasing with VAL, and that VAL largely understood what they meant. However, some reported issues: two participants told VAL ``I want to teach you to \texttt{<x>}'', only for VAL to interpret the utterance as a request to perform a ``teach'' action. One participant also felt that VAL needed explanations at too high a granularity; while we expect that, in repeated uses of the same VAL model, many actions could be re-used to reduce this effect, it is true that VAL starts only with its primitives, and requires explicit explanations for any other actions. It may be worth exploring ways to perform environmental search, or to use language models, to provide initial, non-binding suggestions for how to implement some higher-level actions, in order to reduce the burden on the user.

\subsubsection{Theory of Mind}
Several users reported that they felt unsure of what exactly VAL modeled about the world, and of what operations it could perform on that model. For example, one user asked VAL to edit one of the steps in its current plan, which VAL cannot currently do. Another user was unsure if VAL would face an object after moving to it, and attempted to ask VAL this question; VAL does not have access to this information, and is not designed to respond to questions of this kind, and so gave a confusing response.

These situations highlight the importance of \textit{mutual theory of mind} in conversational agents \citep{mtom}; it is important not only for VAL to model the intent of the human instructor, but for the instructor to be able to model VAL's capabilities and likely responses as well. The more restricted the agent's capabilities, the more important this becomes. We intend to look into ways to augment the displayed knowledge structures, as well as the tutorial process, to  give users a better sense of VAL's cognitive machinations.

\subsubsection{Tedium of Confirmatory Dialogs}
Many users complained that the confirmatory dialogs used to correct and evaluate GPT subroutine outputs were tedious. This was expected to some degree, and while they were necessary for this study, we discuss here some ways we might reduce the number of confirmatory dialogs in actual deployment.

First, as we discussed in Section~\ref{sec:gpt-rates}, the use of \texttt{verbalizeGPT} and \texttt{paraphraseGPT} to check the output of \texttt{mapGPT} was quite successful, and could reduce the number of \texttt{mapGPT} confirmation dialogs by roughly 80\% while hardly increasing the error rate. Using additional GPT subroutine calls to check the output of the other subroutines may lead to similar results across the board, further reducing these dialogs.

Caching user responses may also be useful; as the system stands, users are asked to confirm each time, even if duplicate commands are issued. This happened many times, for example, with the command ``press space''.

In general, confirmatory dialogs present a trade-off: tedium for correctness. However, given the sequential nature of the dialog, one final optimization to consider may be a stricter threshold for requesting confirmation \textit{earlier} in a dialog, while allowing the undo button to become the primary mechanism of error correction \textit{later} in a dialog.

\subsubsection{GPT Model Size}
As we discuss in Section~\ref{sec:gpt-size-perf}, increasing the size of the base GPT model increased performance in five metrics while only decreasing processing speed. In fact, \textit{every} metric saw improvement; those reported earlier were simply the only ones to meet our statistical significance threshold.

However, there are ethical concerns about this method of scaling VAL's performance, the first being the environmental cost of training and using larger and larger language models \citep{danger-parrots}. Additionally, the models we used are not open-source, and the centralization of foundational technology behind the veil of a corporate API offers almost no transparency, both into how the technology works and into how the data sent by users is being used or stored \citep{llm-ethics}.

While we wished to use open-source language models in this work, we determined, in early pilot interactions, the largest models we could feasibly run locally were almost completely ineffective at implementing VAL's GPT submodules.

So, while this work is a proof of concept using the highest-performing existing language models, we now advocate---for parties with the resources to employ them---the use of open-source models \citep{llama,opt,bloom} to power systems like VAL, even if it will take time for their performance to catch up to that demonstrated in this work.

\subsubsection{Generalizability}
The Overcooked-AI environment we chose is commonly employed in interactive machine learning research. On the one hand, it is relatively simple, making it easy for end users to quickly understand and use. On the other hand, it supports a wide range of level designs and cooking-related tasks, which make it possible to challenge users and AI agents with novel situations they have not encountered before. This blend of simplicity with certain kinds of novel complexity makes it uniquely well suited for studying interactive task learning. 
While the environment only possess two primitive actions---moving and pressing the space bar---it features several higher-level tasks (e.g., for preparing and combining ingredients and plating dishes). As shown in Figure~\ref{fig:htn}, a learned ``cook'' action requires the use of 5 lower-level learned and primitive actions. Our user study shows that VAL can scale up to around ten actions. VAL's architecture is, in principle, agnostic to the action space; all primitive and learned actions are considered for mapping to user utterances, regardless of their number, and the most appropriate will ideally be chosen in each case. 


We view the main generalizability concern as a practical, \textit{scalability} concern: as the number of primitive and learned actions grows, the language models responsible for action mapping are likely to experience higher error rates. While this paper demonstrates the fundamentals of the VAL approach, this practical concern will likely have to be addressed in future work and re-examined as available language models improve.

\section{Conclusion}
\label{sec:conclusion}

VAL offers a new angle on the long-standing problem of interactive task learning: using language models in a principled way to enable natural instructional dialog. Using language models allows VAL to comprehend malformed or casual sentences that traditional semantic parsers either cannot, or would require substantial manual engineering to, parse. Furthermore, restricting language model use to a small number of narrow tasks increases interpretability and reliability.

VAL can acquire knowledge incrementally, from few examples, and generalize that knowledge for use in new situations. The natural language interface, confirmatory dialogs, and undo feature increase VAL's usability relative to existing, more brittle systems. Ultimately, we hope that VAL's usability by those with no formal training in programming or AI will help enable a wide variety of people to not only teach VAL new tasks, but also to \textit{learn} tasks from VAL in educational and job training settings.

The user study we conducted supports VAL's usability, but also underscores avenues for future work on VAL: adding condition learning, incorporating additional modalities like demonstration and vision, helping users form a better theory of VAL's mind, reducing GPT subroutine errors, and optimizing the number of confirmatory dialogs are all substantial and important future projects to improve VAL. We hope that the VAL approach continues to evolve and provide a framework for future work on interactive task learning.

\begin{acks}

This work was funded in part by the ARL STRONG Program (Awards \#W911NF2120101, \#W911NF2120126, and \#W911NF2320203) and by the NSF National AI Research Institutes Program (Award \#2112532). The opinions expressed are those of the authors and do not represent the views of the sponsoring agency. We thank Daniel Weitekamp for his insights into how best to frame the work for the CHI community. Lastly, we acknowledge the work of Alexandra Elbakyan.

\end{acks}


\bibliographystyle{ACM-Reference-Format}
\bibliography{val}


\clearpage

\appendix

\section{Examples of All Confirmatory Dialogs}
\label{app:confirm_dialogs}

\begin{figure}[ht]
\begin{minipage}[b][][b]{.8\columnwidth}
\includegraphics[width=\textwidth]{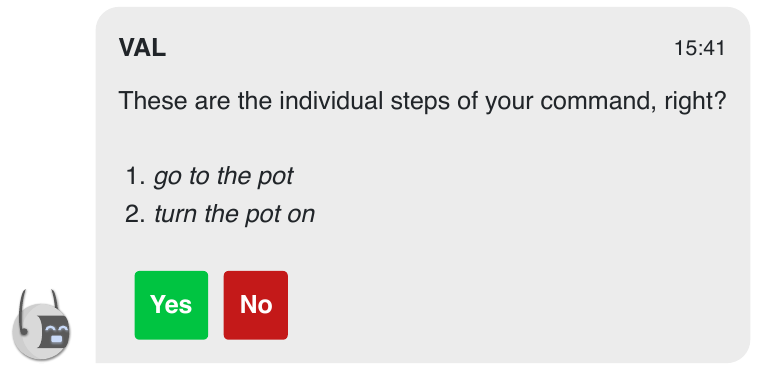}
\caption{\protect\texttt{segmentGPT} confirmation}
\end{minipage}
\begin{minipage}[b][][b]{.8\columnwidth}
\includegraphics[width=\textwidth]{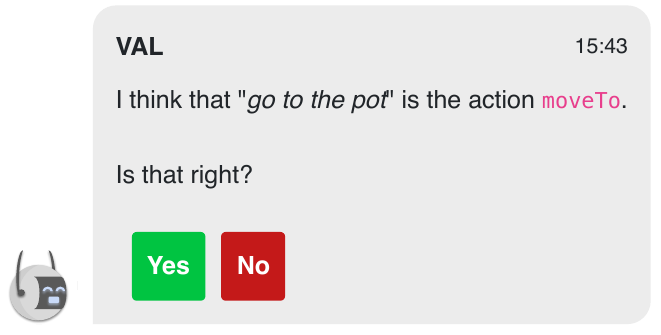}
\caption{\protect\texttt{mapGPT} confirmation}
\end{minipage}
\begin{minipage}[b][][b]{.8\columnwidth}
\includegraphics[width=\textwidth]{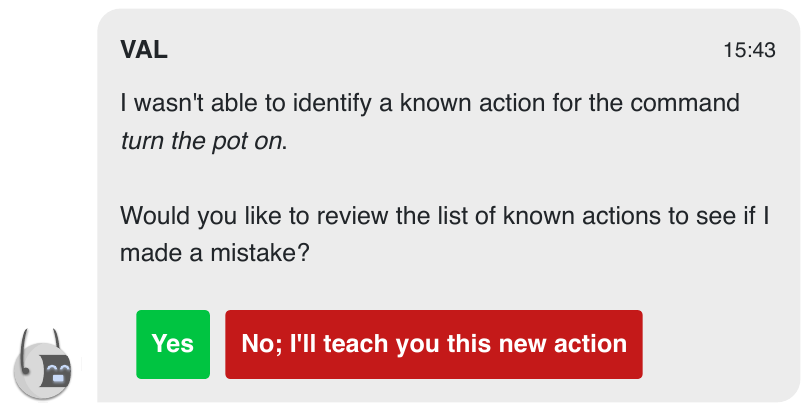}
\caption{\protect\texttt{mapGPT} new-action confirmation}
\end{minipage}
\end{figure}

\begin{figure}[ht]
\begin{minipage}[b][][b]{.8\columnwidth}
\includegraphics[width=\textwidth]{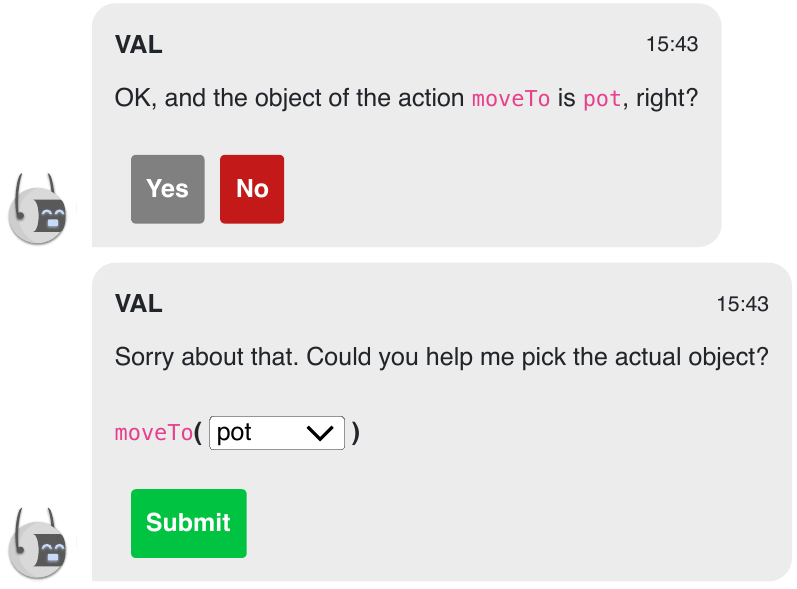}
\caption{\protect\texttt{groundGPT} correction}
\end{minipage}
\begin{minipage}[b][][b]{.8\columnwidth}
\includegraphics[width=\textwidth]{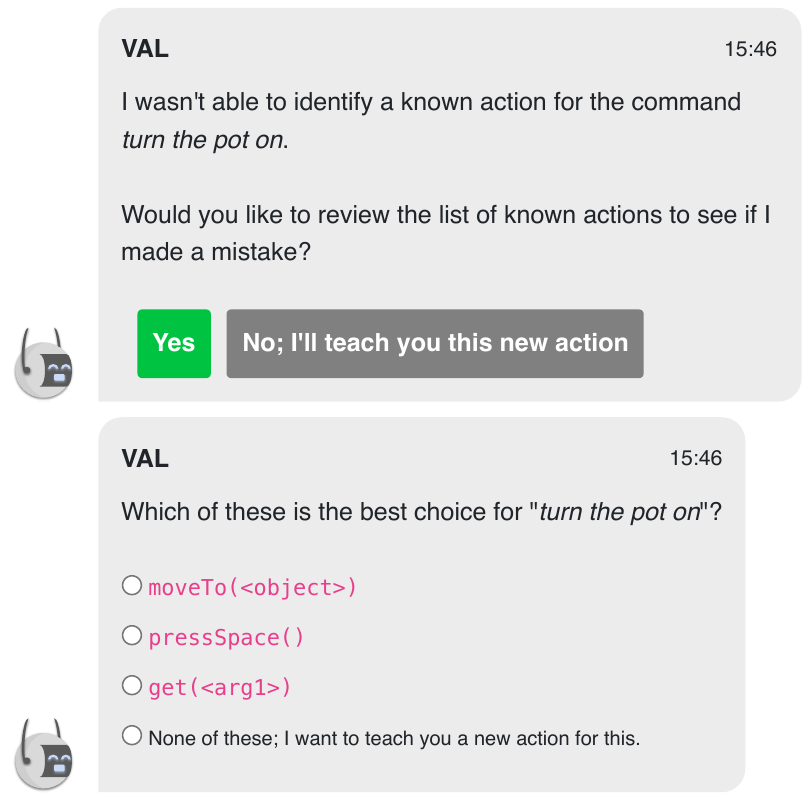}
\caption{\protect\texttt{mapGPT} correction}
\end{minipage}
\begin{minipage}[b][][b]{.8\columnwidth}
\includegraphics[width=\textwidth]{figures/confirm_dialogs/confirm_correct_arg_choice.pdf}
\caption{\protect\texttt{genGPT} correction}
\end{minipage}
\end{figure}

\begin{figure}[ht]
\begin{minipage}[b][][b]{.8\columnwidth}
\includegraphics[width=\textwidth]{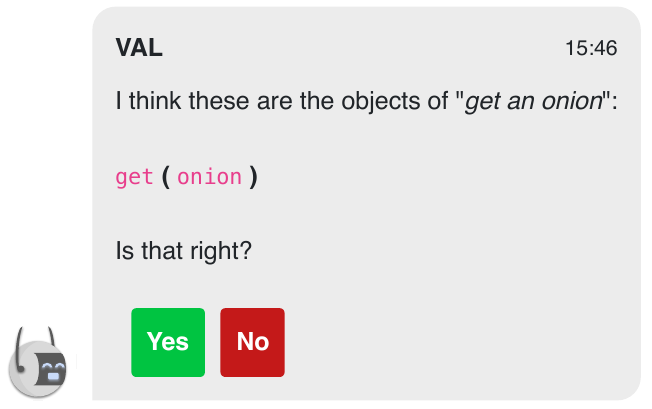}
\caption{\protect\texttt{genGPT} confirmation}
\end{minipage}
\begin{minipage}[b][][b]{.8\columnwidth}
\includegraphics[width=\textwidth]{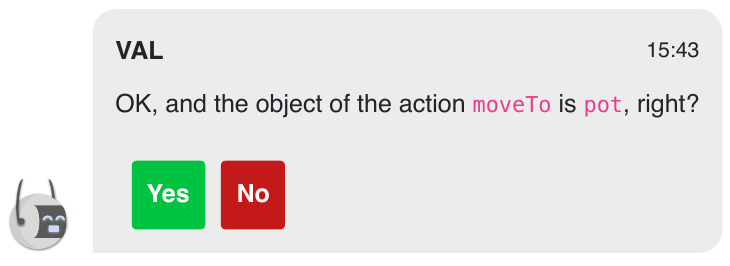}
\caption{\protect\texttt{groundGPT} confirmation}
\end{minipage}
\begin{minipage}[b][][b]{.8\columnwidth}
\includegraphics[width=\textwidth]{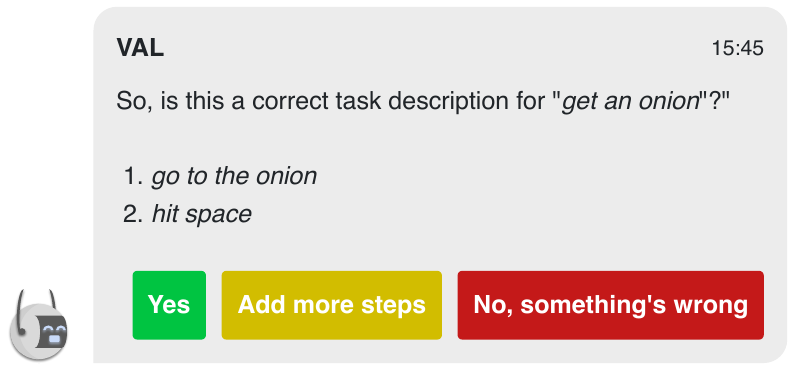}
\caption{Task correctness confirmation}
\end{minipage}
\end{figure}

\clearpage

\section{VAL Participant Survey Form}
\label{app:survey}
\vspace{1in}
\includegraphics[width=\textwidth]{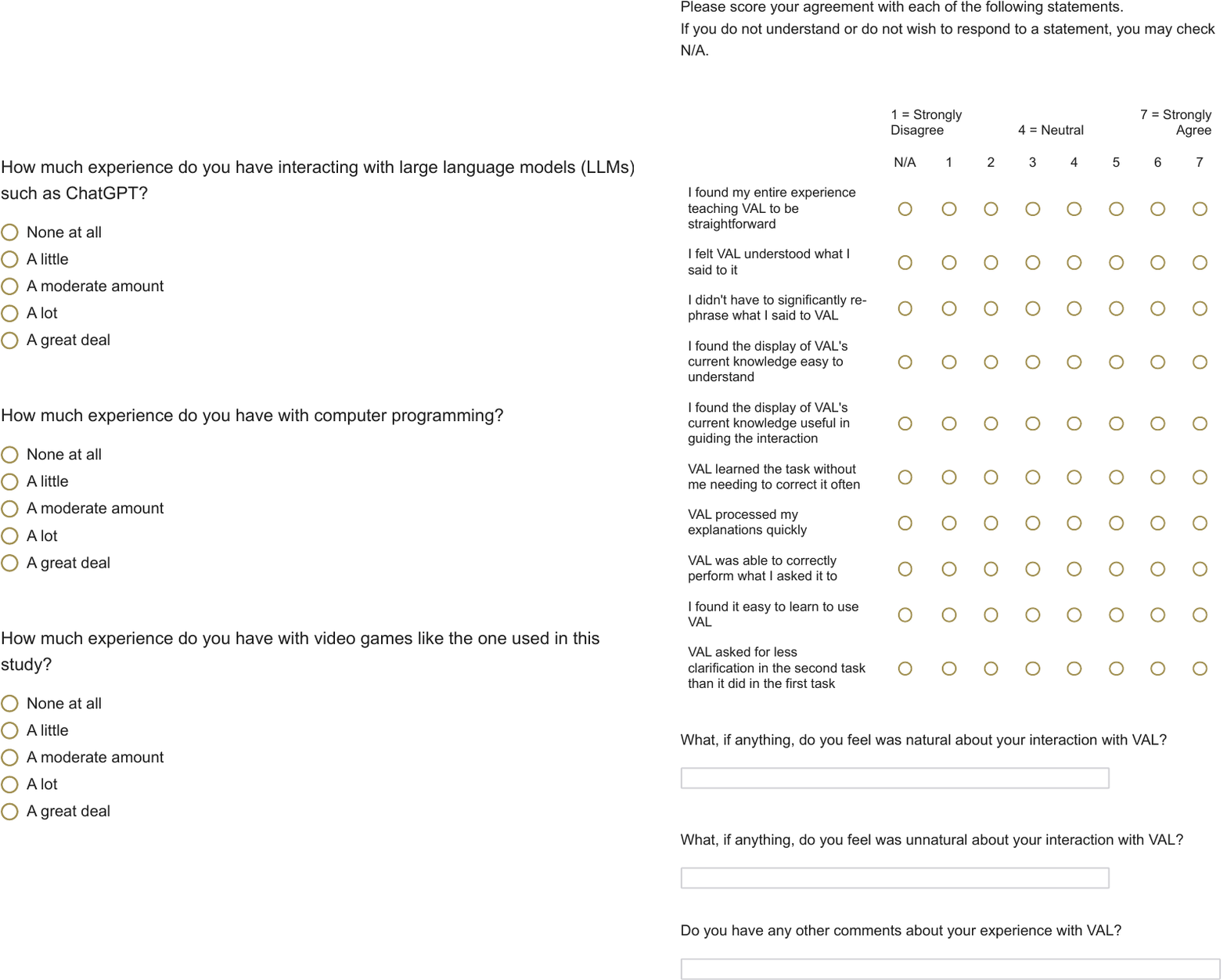}

\clearpage

\section{Prior Experience Survey Question Results}
\label{app:prior_exp_results}

\begin{figure}[ht]
\begin{minipage}[b][][b]{\textwidth}
\includegraphics[width=\textwidth]{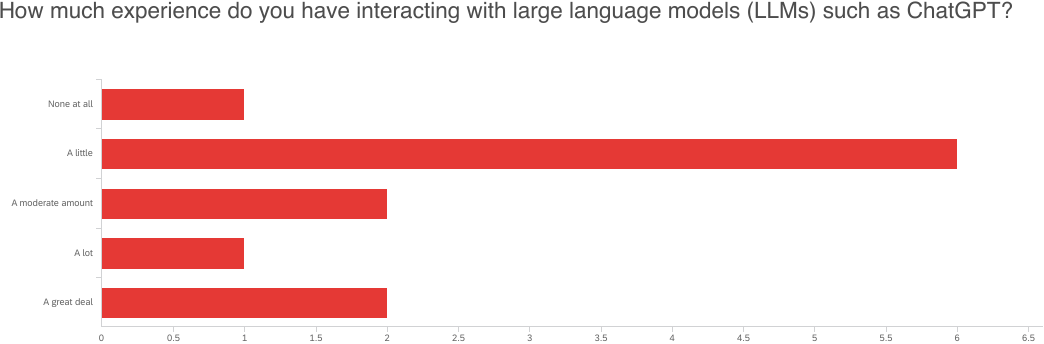}
\end{minipage}\\
\vspace{10mm}
\begin{minipage}[b][][b]{\textwidth}
\includegraphics[width=\textwidth]{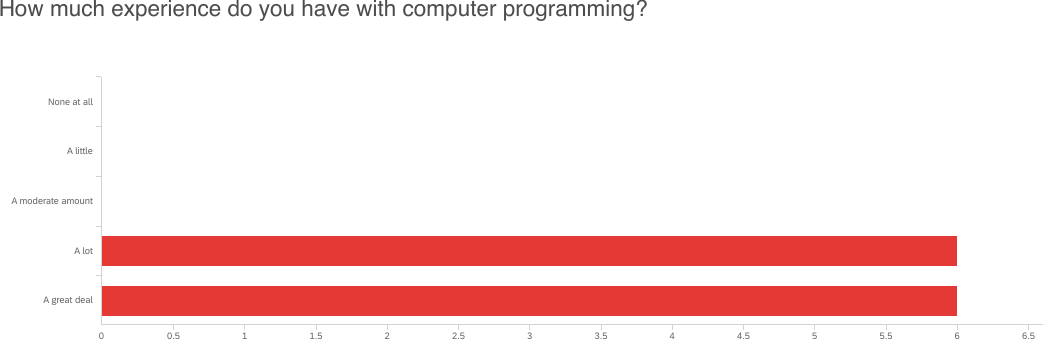}
\end{minipage}\\
\vspace{10mm}
\begin{minipage}[b][][b]{\textwidth}
\includegraphics[width=\textwidth]{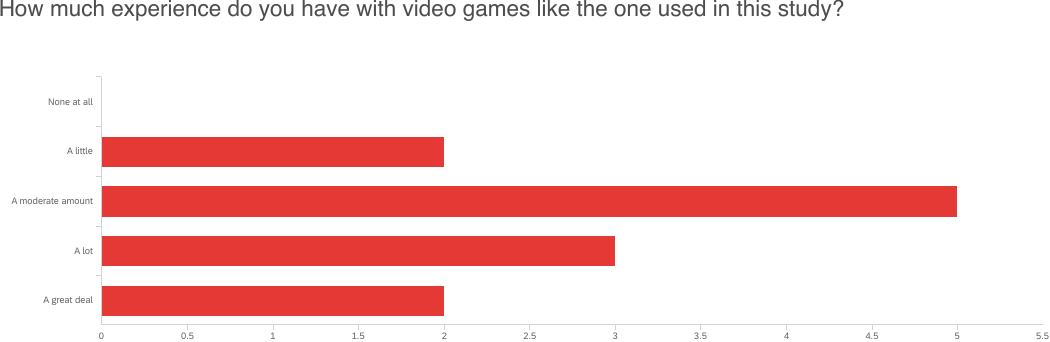}
\end{minipage}
\end{figure}

\clearpage
\twocolumn

\section{Alternative Abstract Using Only 1,000 Most Common Words}
\label{app:upgoer5}

Until now, when people have tried to make computers you can talk with to explain how to do things, the computers have used old-school ways of understanding the words. The old ways they use sometimes can't understand people's words for no good reason. These days, some people use large ``attention is all you need'' approaches to word understanding instead, but those approaches sometimes see things that aren't there, or try to do things they can't do. In this paper, we're going to talk about a new way, where we use an old-school approach to learning, but add in a lot of large-attention pieces who each have very small and focused jobs. The attention pieces help the old-school piece do its job better, and the old-school piece makes sure the attention pieces don't say weird stuff or see things that aren't there. This lets us have a computer that can turn words into an understanding of how to play a game. We asked some people to help us explain to the computer how to play the game, and most of them said it was easy to do.

\begin{figure}[h]
\centering
\includegraphics[width=0.5\textwidth]{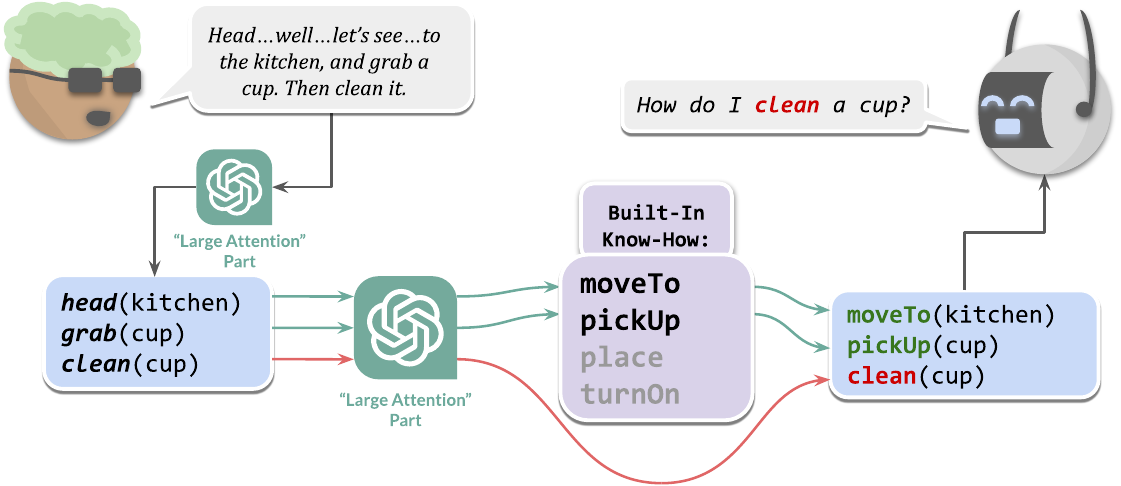}
\caption{\textsf{A drawing showing one step of our computer learning. Plan~\ref{alg:valgorithm} and Part~\ref{sec:gpt_subroutine_arch} do a better job of explaining the full way our computer learns how to do things from talking to people.}}
\Description{A simple picture showing a human being explaining how to do something to our computer, which asks a follow-up question after thinking about the words using several ``large attention'' pieces with their own small, focused jobs.}
\label{fig:upgoer5_teaser}
\vspace{3mm}
\end{figure}



\end{document}